\documentclass[aps,prl,twocolumn,groupedaddress]{revtex4-1}

\usepackage{amsmath}
\usepackage{graphicx}
\usepackage{sistyle}
\usepackage{todonotes}
\usepackage{gensymb}

\usepackage{bm}
\usepackage{natbib}
\usepackage{dsfont}

\usepackage{epsfig}

\begin{document}

\title{Nonlinear interaction between two heralded single photons}

\author{T.~Guerreiro$^1$, A.~Martin$^1$, B.~Sanguinetti$^1$, J.~S.~Pelc$^{2}$, C.~Langrock$^2$, M.~M.~Fejer$^2$, N.~Gisin$^1$, H.~Zbinden$^1$, N.~Sangouard$^1$ \& R.~T.~Thew$^1$}

\email[]{robert.thew@unige.ch}
\homepage[]{www.unige.ch/gap/optics}

\affiliation{$^1$Group of Applied Physics, University of Geneva, 1211 Geneva 4, Switzerland}
\affiliation{$^2$E.L. Ginzton Laboratory, Stanford University, 348 Via Pueblo Mall, Stanford, California 94305, USA}


\begin{abstract}

\end{abstract}


\maketitle


{\bf  Harnessing nonlinearities strong enough to allow two single photons to interact with one another is not only a fascinating challenge but is central to numerous advanced applications in quantum information science~\cite{NielsenChuang}. Currently, all known approaches are extremely challenging although a few have led to experimental realisations with attenuated classical laser light. This has included cross-phase modulation with weak classical light in atomic ensembles~\cite{Kang2013,Chen2006} and optical fibres~\cite{Matsuda2009}, converting incident laser light into a non-classical stream of photon~\cite{Birnbaum2005,Kubanek2008} or Rydberg blockades~\cite{Peyronel2012} as well as all-optical switches with attenuated classical light in various atomic systems~\cite{Dawes2005,Hwang2009,Englund2012,Volz2012,Chen2013}. Here we report the observation of a nonlinear parametric interaction between two true single photons. Single photons are initially generated by heralding one photon from each of two independent spontaneous parametric downconversion sources. The two heralded single photons are subsequently combined in a nonlinear waveguide where they are converted into a single photon with a higher energy. Our approach highlights the potential for quantum nonlinear optics with integrated devices, and as the photons are at telecom wavelengths, it is well adapted to applications in quantum communication~\cite{Gisin2007}.}

Observing nonlinear processes down to the quantum regime is a long sought goal for quantum information science~\cite{NielsenChuang} as well as a fascinating concept in terms of fundamental physics, first being raised in the seminal work of Heisenberg and Euler~\cite{Heisenberg1936}. It is only in recent years that materials and technologies have advanced to the point where one can probe this quantum nonlinear domain. Experimental efforts have focused on nonlinear optical effects in atomic ensembles~\cite{Kang2013,Chen2006,Peyronel2012,Dawes2005,Chen2013} or in coupled cavity-single-atom systems~\cite{Thompson1992,Brune1996,Birnbaum2005,Kubanek2008,Hwang2009,Englund2012,Volz2012}. These atomic systems naturally operate with very narrow bandwidths and at specific wavelengths, typically in the visible regime. A grand challenge is to realise photon-photon interactions in materials that are less restrictive in terms of bandwidths and wavelengths. Of particular interest are photons at telecommunication wavelengths as these provide the wiring, the flying qubits, for myriad applications in quantum communication~\cite{Gisin2007,Sangouard2012}. A further challenge is to not only realise photon-photon interactions in a material that is less restrictive in terms of bandwidth, but ideally compatible with telecom wavelength photons and operating at room temperature.

\begin{figure}[!h]
\centering
\includegraphics[width=1\columnwidth]{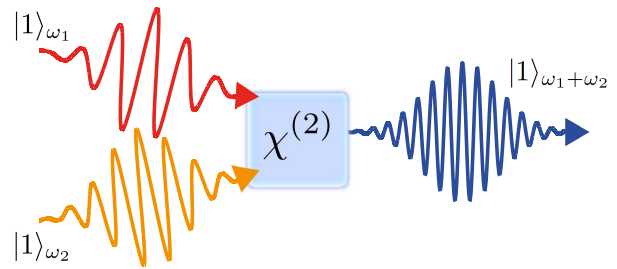}
\caption{Concept: Two single photons are sent to a medium with $ \chi^{(2)} $ nonlinearity and interact, generating a third photon carrying the sum of the energies and momenta of the input fields.}
\label{fig1}
\end{figure}

We have taken an approach that exploits a parametric process in a nonlinear crystal~\cite{Tanzilli2012, Kim2001, Dayan2005, Giorgi2003, Vandevender2004, Thew2008}. The efficiency of nonlinear optical materials is constantly increasing and by benefiting from their inherently large bandwidth to work with pulsed systems at high repetition rates, important experimental results have been obtained in the context of quantum nonlinear optics with parametric processes~\cite{Hubel2010, Shalm2013, Langford2011, Guerreiro2013}, notably the direct generation of photon triplets from the spontaneous conversion of a single photon into a photon pair~\cite{Hubel2010,Shalm2013}. Here we show for the first time a nonlinear interaction between two true independent single photons (Fock states) via sum frequency generation (SFG), as conceptually depicted in \figurename{~\ref{fig1}}. This experiment cannot be seen as the time reversal process of the experiment described in Refs.~\cite{Hubel2010,Shalm2013} as our photons were generated in independent sources and therefore have uncorrelated spectra. This makes sum frequency generation much more challenging but at the same time, it offers unique opportunities for example to herald entangled photon pairs remotely and to perform quantum key distribution where the security is independent of the internal workings of the devices used to generate the secret key~\cite{Sangouard2011}.

\begin{figure*}[!ht]
\centering
\includegraphics[width=1.42\columnwidth]{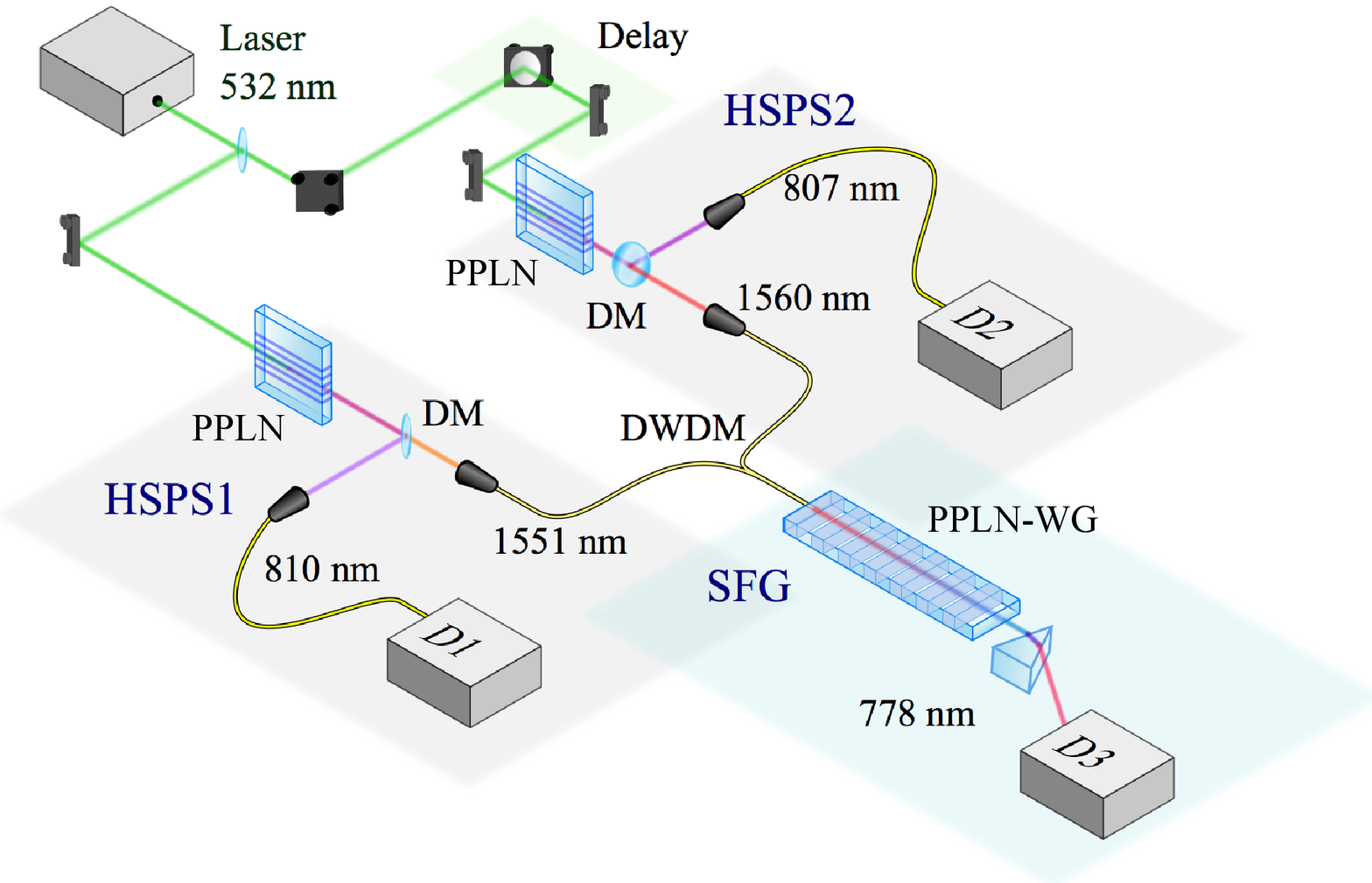}
\caption{Experimental setup. A mode-locked laser [TimeBandwidth] generates \SI{10}{ps} pulses at \SI{532}{nm} with a repetition rate of \SI{430}{MHz} and is used to pump the two heralded single photon sources (HSPS1 \&HSPS2)  based on periodically poled lithium niobate (PPLN) nonlinear crystals. The generated photons are deterministically separated by dichroic mirrors (DM), collimated, and then collected into single-mode optical fibres. Diffraction gratings (not shown) are employed to filter the heralding photons (810,\,807\,nm) down to $\sim$\,\SI{0.3}{nm}. In this configuration the telecom photons are projected onto a spectral mode that is matched to the acceptance bandwidth of the SFG process, which was measured to be \SI{0.27}{nm}. The two fibre coupled telecom photons are combined via a dense wavelength division multiplexer (DWDM) and directed to a \SI{4.5}{cm}-long fibre-pigtailed Type-$ 0 $ PPLN waveguide~\cite{Parameswaran2002}. The unconverted photons are deterministically separated from the SFG photons by a prism (not shown)and the upconverted light is sent to a single photon detector D3. We record threefold coincidences between detectors D1, D2 and D3 using a time-to-digital converter TDC [QuTools]. The overall  SFG conversion efficiency is $ 1.56 \times 10^{-8} $, including the coupling of the fibre pigtail which is \SI{70}{\%}. }
\label{fig2}
\end{figure*}

The experiment has three distinct parts: the generation of two independent single photons by two heralded single photon sources (HSPS); their parametric interaction in a nonlinear waveguide, and finally the detection of the resulting single photon of higher energy. A schematic of the setup is shown in figure \figurename{~\ref{fig2}}. 

The single photons are generated via spontaneous parametric downconversion (SPDC) in two independent sources. HSPS1 and HSPS2 generate pairs at \SI{807}{nm} - \SI{1560}{nm} and \SI{810}{nm} - \SI{1551}{nm}, respectively. By ensuring that the probability of creating a single pair in each source is much smaller than one, the detection of  the visible (807, 810\,nm) photons heralds the creation of two independent single telecom wavelength (1560, 1551\,nm) photons. All the photons are coupled into single-mode fibres with efficiencies $\sim$\,50\,\%~\cite{Guerreiro2013b}. Importantly the heralding photons are filtered such that the bandwidth of telecom photons is matched to the acceptance bandwidth of the SFG process, c.f. below. To verify the single photon nature of these sources, we measured the conditional second-order autocorrelation functions $ g^{(2)}_{1}(0) = 0.030 $ and $ g^{(2)}_{2}(0) = 0.036 $, for HSPS1 and HSPS2 respectively.

\begin{figure}[!htp]
\includegraphics[width=0.5\textwidth]{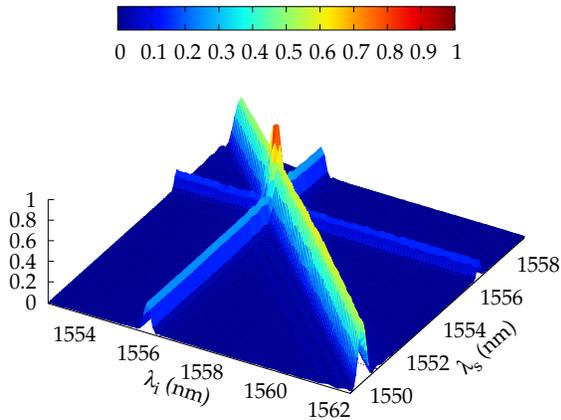}
\caption{Intensity plot of the classical SFG efficiency as a function of the wavelength of two input fields. The diagonal trace represents SFG between the input fields, while the horizontal and vertical traces signal the contributions from second harmonic generation (SHG) of each individual input field. Note that the difference in hight between the SFG and SHG signals is of about a factor of 4, as expected. It is also possible to see the oscillations from the $ \rm{sinc}^{2} $ structure of the phasematching on the leading edge.}
\label{fig3}
\end{figure}

\begin{figure*}[ht!]
\begin{tabular}{cc}
a) \includegraphics[width=\columnwidth]{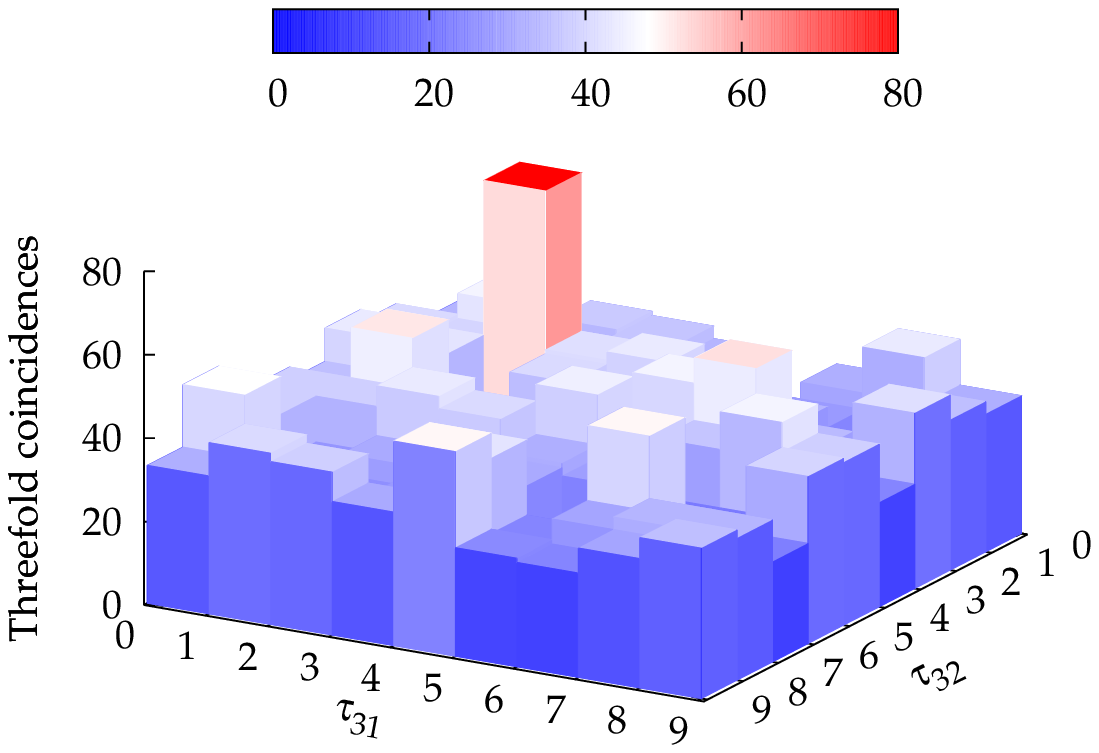}&
b) \includegraphics[width=\columnwidth]{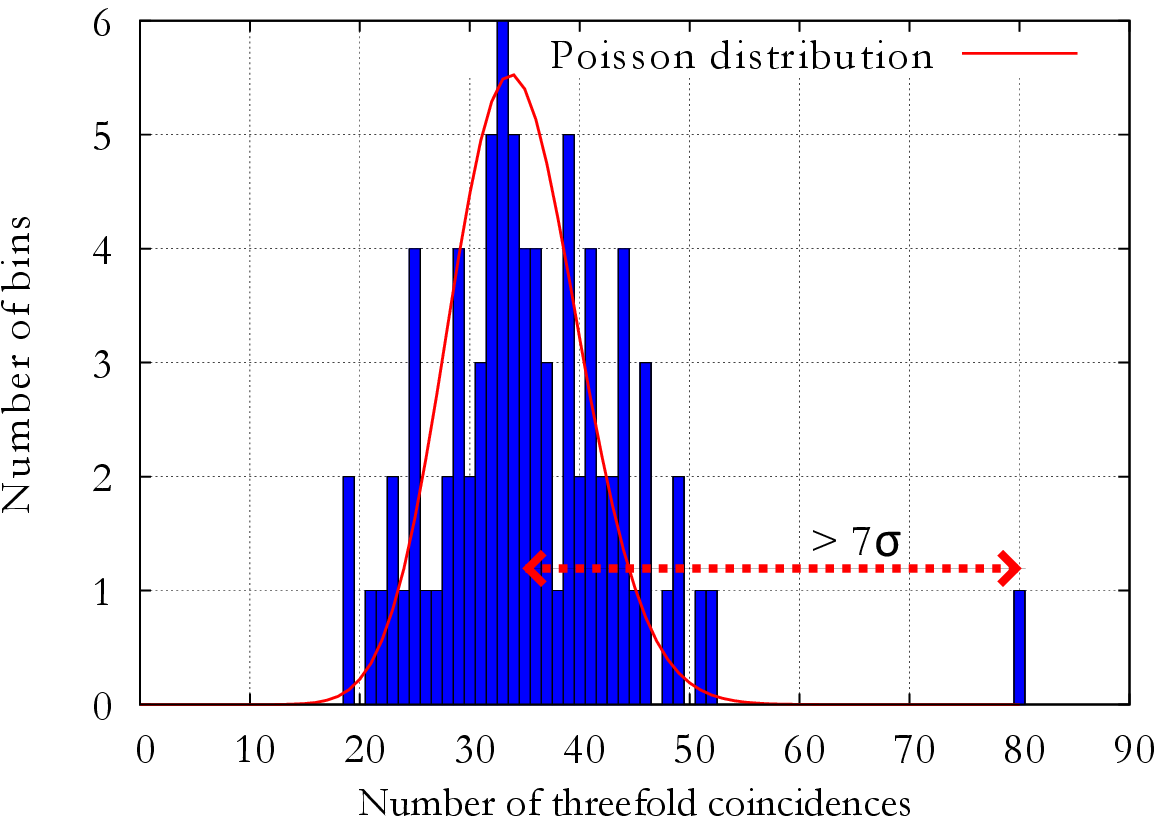}
\end{tabular}
\caption{\label{fig4}\textbf{a)} A clear signature of the photon-photon interaction can be seen to emerge from the background noise in the time-of-arrival threefold coincidence histogram. The axis labeled $ \tau_{31} $ shows the delay between detectors D1 and D3, while $ \tau_{32} $ shows the delay between D2 and D3. Each pixel is composed of \SI{2.3}{ns} bins, defined by the laser repetition rate. \textbf{b)} The histogram of three-fold coincidence counts shows a Poissonian distribution for the noise with a mean value of 35. The single peak at 80 corresponds to the true threefold coincidences and exceeds the mean value by over $ 7$ standard deviations ($7\,\sigma $).}
\end{figure*}

The two fibre coupled telecom photons are then directed to a fibre-pigtailed PPLN waveguide that is quasi-phase matched to perform the SFG process $ \SI{1560}{nm}~+~\SI{1551}{nm} ~\rightarrow~\SI{778}{nm}$. \figurename{~\ref{fig3}} shows the results of a classical measurement of the phasematching conditions of the waveguide. The diagonal ridge corresponds to the SFG process, while the horizontal and vertical ridges result from the second harmonic generation (SHG) of each individual field. Following~\cite{Guerreiro2013} and taking into account the acceptance bandwidth of the waveguide and the bandwidth of the interacting photons, which are measured to be \SI{0.27}{nm}, we determine a system conversion efficiency of $ 1.56~\times~10^{-8} $.

Finally, the detection scheme consists of three single photon detectors D1, D2 and D3 based on Si avalanche photodiodes. Detector D3 [Picoquant:  $\tau$-SPAD-20] operates in free-running mode with an efficiency of $ 60 \% $ at \SI{778}{nm} and darkcount rate of 3.5\,s$^{-1}$. Detectors D1 and D2 [Excelitas diodes with custom electronics] are gated (18\,ns) devices with an efficiency of 60\% and a darkcount probability of $10^{-6}/$\,ns~\cite{Lunghi2012}.  All the detection events from D1, D2 and D3 are recorded by a time-to-digital converter (TDC).

Before running the nonlinear interaction measurement, it is necessary to make sure the interacting photons arrive at the same time into the nonlinear waveguide. To guarantee this, we seed HSPS2 with a continuous wave (cw) laser at \SI{810}{nm}, producing a pulsed coherent state at \SI{1551}{nm} by difference frequency generation (DFG)~\cite{Bruno2013} to improve the signal to noise level. We record twofold coincidences between D3 and D2 as we scan the delay line placed before HSPS2. This allows for the temporal alignment of the interacting photons with picosecond resolution and also allows us to determine the exact position of the three-fold coincidence peak in \figurename{~\ref{fig4}}\textbf{a}.   

Once the two photons are temporally aligned, we can remove the seed laser and use the two sources in the heralded single photon configuration. We then proceed to record threefold coincidences between D1, D2 and D3, where the appearance of a peak in the threefold detection time histogram signals the correlated generation of triplets of photons and hence that the interaction between the two independent telecom photons has taken place. We denote the delay between D1 and D3 as $ \tau_{31} $, and the delay between D2 and D3 as $ \tau_{32} $.


The threefold coincidences between D1, D2 and D3 are shown in the time-of-arrival histogram in \figurename{~\ref{fig4}}\textbf{a}. Each bin of this histogram corresponds to an acquisition window of \SI{2.3}{ns} for each detector matching the repetition rate of the pump laser. We integrate for \SI{260}{hours} and observe a well defined coincidence peak exactly where it is expected. Moreover, \figurename{~\ref{fig4}}\textbf{b} shows the histogram of three-fold coincidence counts. One sees a Poissonian distribution for the background noise with a mean value of 35. The background noise is dominated by the detection of photons at D1 and D2 in coincidence with darkcounts (3.5\,s$^{-1}$) from detector D3. The three photon signature is the single pixel containing 80 counts, which has a statistical significance of over $ 7$ standard deviations with respect to the background. Furthermore, given the Poisson distribution with a mean value of 35, the probability of having a pixel with 80 accidental counts is of the order of $10^{-11}$.  

Our theoretical model of the system takes into account the source emission probability, the losses of the setup, the SFG conversion probability and the detector efficiency and noise levels. We estimated a rate of 0.40 three-fold coincidences versus a rate of 0.20 threefold noise events, per hour, while the observed values are 0.31 and 0.13 coincidences respectively. The measured values are both slightly reduced due to source alignement drifting over the long integration time. Nonetheless the signal to noise ratio is in good agreement with our predictions.


These results demonstrate the first interaction between two single photons. Despite the challenging nature of the current experiment, significant improvements in the signal to noise ratio could be achieved with lower noise detectors and improved photon coupling efficiencies. Nonetheless, the reported nonlinearity already offers  promising perspectives in quantum information~\cite{Sangouard2011}. For example, the implemention of quantum key distribution, where the secrecy is independent of the internal workings of the devices that are used to create the key (device-independent QKD). An important aspect in this framework is that the combination of entangled photon pairs at telecom wavelengths and the SFG process allows for maximally entangled photon pairs to be created at a distance while being heralded through the detection of converted photons~\cite{Sangouard2011}. The presented demonstration of photon-photon interaction is also promising from a fundamental perspective, for instance unambiguously excluding local hidden variable models of entanglement in a loophole-free Bell-type experiment~\cite{Cabello2012}, opening the way for investigating novel quantum correlations~\cite{Banaszek1997} and providing a platform for studying exotic states of light and quantum optical solitons~\cite{Kheruntsyan1998}. 

Looking further ahead, higher efficiency nonlinear interactions could be obtained by using tight spatial confinement of the optical modes~\cite{Kurimura2006}, from the use of highly nonlinear organic materials~\cite{Jazbinsek2008} or by exploiting weak measurements based on pre- and post-selected states, as pointed out in Ref.~\cite{Feizpour2011}. Aside from these exciting perspectives, we believe that our demonstration of an interaction between two independent single photons will strongly stimulate research in nonlinear optics in the quantum regime.

\begin{acknowledgments}
The authors would like to thank E. Pomarico for useful discussion. This work was partially supported by the EU Projects SIQS and Chist-Era:~Qscale and DIQIP, as well as the Swiss NCCR QSIT and US AFOSR (FA9550-12-1-0110).
\end{acknowledgments}

\bibliography{SFG_Bibliogrophy}

\end{document}